\documentclass[prd,preprint,eqsecnum,nofootinbib,amsmath,amssymb,tightenlines,superscriptaddress,dvips]{revtex4}
\usepackage{graphicx}
\begin{document}

\title{Quarkonium spectral functions with complex potential}


\author{Chuan Miao}
\affiliation{Physics Department, Brookhaven National Laboratory, Upton NY 11973 USA}
\affiliation{Institute of Nuclear Physics,Johannes Gutenberg-Universit\"at Mainz,
Johann-Joachim-Becher-Weg 45, D-55099 Mainz, Germany  }

\author{\'Agnes M\'ocsy}
\affiliation{Pratt Institute, Department of Math and Science, Brooklyn, NY 11205, USA}

\author{P\'eter Petreczky}
\affiliation{Physics Department, Brookhaven National Laboratory, Upton NY 11973 USA}

\begin{abstract}
We study quarkonium spectral functions at high temperatures using
potential model with complex potential. The real part of the potential
is constrained by the lattice QCD data on static quark anti-quark correlation functions,
while the imaginary part of the potential is taken from perturbative calculations.
We find that the imaginary part of the potential has significant effect on quarkonium spectral
functions, in particular, it leas to the dissolution of the 1S charmonium and excited bottomonium
states at temperatures about $250$MeV and melting of the ground state bottomonium at
temperatures slightly above $450$MeV.
\end{abstract}

\date{\today}


\maketitle

\section{Introduction}
\label{intro}
It is known from lattice QCD that strongly interacting  matter undergoes
a transition to a deconfined phase also called quark gluon plasma (QGP)
characterized by chiral symmetry restoration and color screening \cite{me1}.
Quarkonia have been suggested as experimental signature of deconfinement
by Matsui and Satz \cite{Matsui:1986dk}. 
Namely, it has been argued that color screening in a deconfined QCD
medium will suppress the existence of quarkonium states, signaling the
formation of QGP in heavy-ion collisions. Although
this idea was proposed a long time ago, attempts to calculate quarkonium
spectral functions from first principle QCD  have been made only relatively recently.
The idea behind these attempts is to calculate the correlation function of the corresponding
meson operators in Euclidean time in lattice QCD and to reconstruct the quarkonium 
spectral functions using the Maximum Entropy Method (MEM)
\cite{Umeda:2002vr,Asakawa:2003re,Karsch:2002wv,Datta:2003ww,Datta:2006ua,Jakovac:2006sf,Aarts:2007pk}. 
It turned
out that the spectral functions of S-wave quarkonium in the deconfined phase 
are similar to those in the confined phase, and this fact was interpreted as survival of
the ground state quarkonium  in the high temperature region. However, as this was pointed out
in Ref. \cite{Jakovac:2006sf} details of the spectral functions cannot be resolved at high
temperatures due to limited extent of the Euclidean time direction (see Ref. \cite{my_qgp4}
for recent review on this issue). Furthermore, lattice artifacts
at small separations (large momenta) further complicate the extraction of the spectral
functions \cite{lat_spf}. On the other hand, the temperature 
dependence of the quarkonium correlation functions
can be studied reliably in lattice QCD. It was found that in the pseudo-scalar and vector
channels corresponding to ground state quarkonia the temperature dependence of quarkonium
correlation functions is either very small or moderate \cite{Datta:2003ww,Jakovac:2006sf,Aarts:2007pk}.
At the same time large temperature dependence in the scalar and axial-vector channels was observed. 
These findings were interpreted as survival of the ground state 
quarkonium and the dissolution of the excited
P-states, which fitted well into the expected sequential melting pattern. However, 
a detailed study of the temperature dependence of the quarkonium correlators revealed that
almost the entire temperature dependence of the quarkonium correlators is due to the zero modes
not to the dissolution of the bound states \cite{Petreczky:2008px}, thus challenging the above
interpretation.

With progress in lattice calculations of the correlation functions of static
quark anti-quark pairs and better understanding of the color screening phenomenon
there has been a renewed interest in potential models at finite temperature
\cite{Digal:2001ue,Wong:2004zr,Mocsy:2005qw,Alberico:2006vw,Cabrera:2006wh,Mocsy:2007yj,Mocsy:2007jz}.
The basic idea of this approach is to assume that medium effects on quarkonium properties can
be incorporated in a temperature dependent potential and use lattice QCD to constrain its form.  
It is not clear whether and under which circumstances medium effects can be characterized by
a temperature dependent potential. Fortunately, the effective field theory approach 
to heavy quark bound states provides an answer \cite{Brambilla:2008cx}. Heavy quark bound states 
at zero temperature are characterized by three distinct energy scales $m\gg mv \gg mv^2$ 
corresponding to the heavy quark mass, the typical momenta inside the bound state, and
the typical binding energy respectively. The heavy quark velocity $v$ can be treated as a small
expansion parameter and furthermore, in the weak coupling regime $v \sim \alpha_s$. This allows
construction of sequence of the effective field theories 
called NRQCD and pNRQCD by integrating out the scales $m$ and $m v$ \cite{Brambilla:1999xf} 
(see Ref. \cite{Brambilla:2004jw} for a detailed
review on this subject). The degrees of freedom in pNRQCD are the singlet and the octet meson fields
composed of the heavy quark and anti-quark and interacting with ultra-soft gluons, i.e. gluons
on energy scale $m v^2$. The singlet and octet potentials appear as the parameters of the effective
field theory Lagrangian and therefore can be defined at any order in perturbation theory or 
even non-perturbatively \cite{Brambilla:1999xf}. The effective field theory approach can be 
generalized to non-zero temperature, but the presence of additional thermal scales $T$, $m_D\sim g T$
and $g^2 T$ ( with $g$ being the gauge coupling, $g^2=4 \pi \alpha_s$) 
make the analysis more complicated \cite{Brambilla:2008cx}. In the case when the binding energy
is smaller than the temperature scales the potential receives thermal corrections which
have both real and imaginary parts and one gets different versions of thermal pNRQCD.
The precise form of the real and imaginary parts of the
thermal corrections depends on the relation of the scales $1/r \sim m v$, $T$ and $g T$ 
\cite{Brambilla:2008cx}.\footnote{The presence of the imaginary part in the 
finite temperature potential was first pointed out in Ref. \cite{Laine:2006ns} 
by analyzing Wilson loops in the hard thermal loop
approximation. In the effective field theory framework it can be shown that this analysis
corresponds to calculation of the potential in the distance regime $1/r \sim g T$.}

In the past quarkonium spectral functions in QGP have been calculated in Refs. 
\cite{Mocsy:2005qw,Cabrera:2006wh,Mocsy:2007yj,Mocsy:2007jz} using potential model
with real potentials.
The aim of the present paper is to perform a calculation of quarkonium spectral functions
in the deconfined phase using a potential model inspired by thermal pNRQCD, i.e. using
a complex potential. The rest of this paper is organized as follows: In section \ref{sec:f1}
we review what is known about the potential at finite temperature from lattice QCD and
perturbation theory. In section \ref{sec:num} we present our numerical results for
the quarkonium spectral functions.
In section \ref{sec:corr} we discuss the temperature dependence of the corresponding
Euclidean time quarkonium correlation functions.
Finally, section \ref{sec:concl} contains our
conclusions. 

\section{The potential at finite temperature}
\label{sec:f1}
\begin{figure}[ht]
\includegraphics[width=8cm]{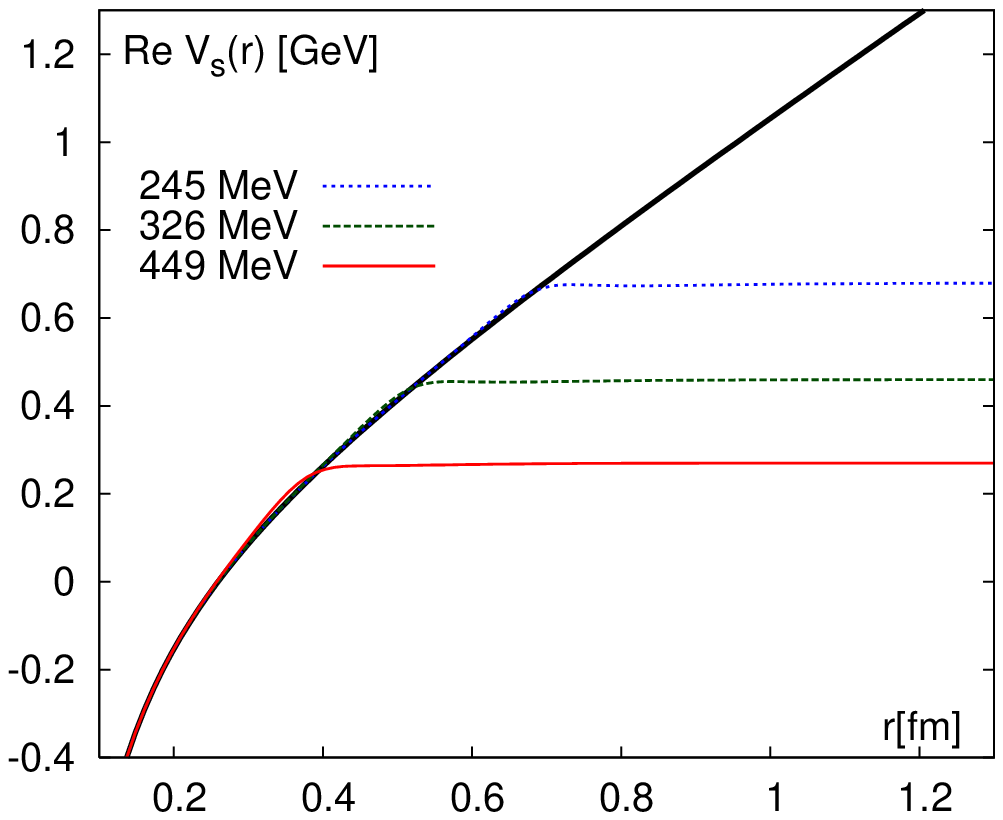}\hspace*{-0.3cm}
\includegraphics[width=8cm]{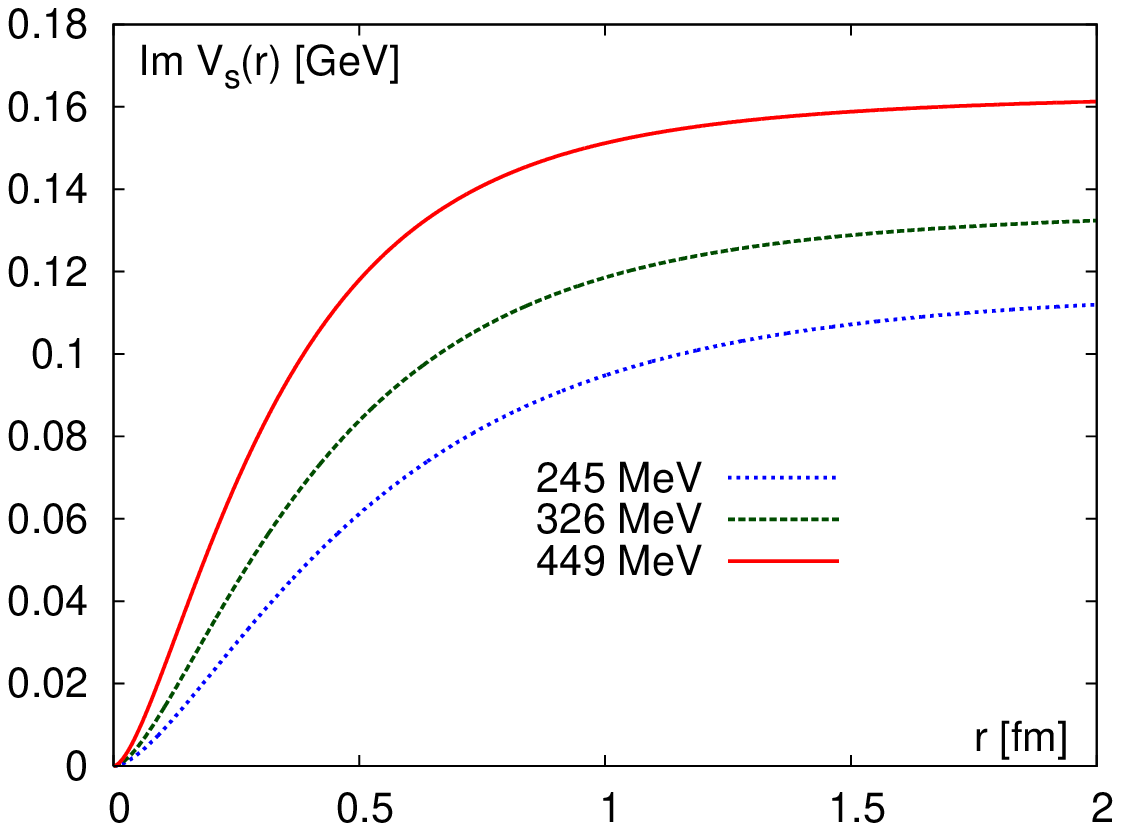}
\vspace*{-0.3cm}
\caption[]{The real (left) and the imaginary part (right) of the potential
used in our analysis.
}
\label{fig:pot}
\vspace*{-0.1cm}
\end{figure}
We are interested in studying quarkonium spectral functions in the
temperature region where the medium significantly changes the properties of
the bound states, i.e. the bound states are close to being dissolved.
This means that the binding energies are significantly reduced from their vacuum values
and eventually vanish  at some temperature called the point of zero binding.
Thus, in this situation the binding energy is the smallest scale in the problem
and all the medium effects can be incorporated in the potential. If we start from
the pNRQCD and neglect the octet-singlet interactions the 
dynamics of the singlet quark anti-quark field is given by the free field
equation which is the form of the Schrodinger equation (e.g. see discussions in
Ref. \cite{Brambilla:2004jw}). Then the calculation of the correlation function
of singlet fields in real time or equivalently the quarkonium spectral function
in non-relativistic limit is reduced to calculating the Green function of the 
Schr\"odinger equation \cite{Burnier:2007qm}
\begin{eqnarray}
&
\displaystyle
\left [ -\frac{1}{m} \vec{\nabla}^2+V(r,T)-E \right ] G^{nr}(\vec{r},\vec{r'},E,T) = \delta^3(r-r')
\nonumber\\[2mm]
&
\displaystyle
\sigma(\omega,T)=\frac{6}{\pi} {\rm Im} G^{nr}(\vec{r},\vec{r'},E)|_{\vec{r}=\vec{
r'}=0}\,~~~~ E=\omega-2 m.
\label{schroedinger}
\end{eqnarray}
The potential entering the above equation is complex \cite{Brambilla:2008cx,Laine:2006ns}.
Its form is quite complicated even in the weakly coupled regime as the form of the
thermal corrections depends on the relation of the scales $1/r$, $T$, $m_D \sim g T$.
Furthermore, in the interesting temperature regime the separation of the above scales does not
hold and the potential is effected by the non-perturbative scales $g^2 T,~\Lambda_{QCD}$. Therefore we 
have to rely on the lattice calculations to constrain the form of the potential.

The interaction of a static quark and anti-quark ($Q\bar Q$) at finite temperature 
is usually studied on 
the lattice in terms of the so-called singlet correlation functions 
$G_1(r,T)={\rm Tr}\langle W(r) W^{\dagger}(0) \rangle$ with $W$ being the temporal Wilson line.
It has been calculated in pure gauge theory \cite{okacz02,Digal:2003jc}, 
in 3-flavor QCD \cite{Petreczky:2004pz},
in 2-flavor QCD \cite{Kaczmarek:2005ui} and more recently also in 2+1 flavor QCD 
\cite{Petreczky:2010yn,Kaczmarek:2007pb,Petrov:2007ug}.
Since the expression for $G_1(r,T)$ is not gauge invariant 
the calculations are performed in Coulomb gauge.
The singlet correlator can be viewed as the correlation function of color singlet static meson field at 
Euclidean time separation $\tau=1/T$. Therefore it is related to the static energy of a $Q \bar Q$ 
pair at
finite temperature (see discussion in Ref. \cite{my_qgp4}), though this relation can be quite complicated.
The other type of static quark anti-quark correlation function that is calculated on the lattice 
is the Polyakov
loop correlator $G(r,T)=\langle {\rm Tr} W(r) {\rm Tr} W^{\dagger}(0) \rangle$ which is gauge invariant.
Unfortunately, its temperature dependence is much more complicated, because it is sensitive 
to the excited  states \cite{Petreczky:2010yn}
\footnote{In leading order of perturbation theory the Polyakov 
loop correlator can be written as the thermal average
of color singlet and octet contribution \cite{McLerran:1981pb,Nadkarni:1986cz}. 
Recently this has been proved
at next-to-leading order in Ref. \cite{Brambilla:2010xn}, 
where it was also shown how to define rigorously the singlet
and octet contribution using pNRQCD. 
In the low temperature limit it can be shown that all excited states contribute
to the Polyakov loop correlator with 
the same weight as the ground state \cite{Jahn:2004qr}, 
while for the singlet correlator the contribution of 
the excited states is suppressed \cite{Bazavov:2008rw}.}.
We also note the  properties of the singlet correlator that are discussed below are not specific
to the Coulomb gauge. In fact, the singlet correlation function defined through periodic Wilson loops at
finite temperature shares all the properties of the 
Coulomb gauge singlet correlator \cite{Bazavov:2008rw}.

The logarithm of the singlet correlator defines the 
so-called singlet free energy $F_1(r,T)/T=-\ln G_1(r,T)$.
For the further discussion it is useful to briefly discuss the properties of the singlet free energy.
At short distances the singlet free energy coincides with the zero temperature potential. 
In the deconfined region, medium effects
become significant at distances $r_o=0.4{\rm fm}/(T/T_{dec})$ 
with $T_{dec}$ being the deconfinement temperature 
\cite{okacz02}. 
At distances $r_e=r T \ge 0.8$ the singlet free energy is 
exponentially screened, i.e. $F_1(r,T)-F_{\infty}(T)$ 
falls off exponentially \cite{Petreczky:2010yn}. 
The constant $F_{\infty}(T)$ is twice the free energy of an isolated static quark and it
is related to the gauge invariant renormalized Polyakov loop expectation value 
$L_{ren}(T)=\exp(-F_{\infty}(T)/(2 T))$.
We expect that the real part of the  potential Re$V(r,T)$ entering in Eq. (\ref{schroedinger}) 
will share the same qualitative features as the singlet free energy, 
in particular medium effects will be significant at some distance $r_o \le r_{med}(T) \le r_e$.
Its value will  depend on the choice of  $r_{med}(T)$. In our analysis we choose 
$r_{med}(T)=r_e$ and assume that for distances smaller than $r_{med}$ the real part of
the potential coincides with the zero temperature potential, while for $r>r_{med}(T)$ it
is exponentially screened \footnote{Note that the nature of this screening is strongly
non-perturbative \cite{Kajantie:1997pd,Karsch:1998tx}.}.
Furthermore, we use a smooth interpolation between these two
regimes as described in Ref. \cite{Mocsy:2007jz}. This procedure also determines the asymptotic
value of the potential $V_{\infty}(T)$ which turns out to be close to value of the internal 
energy of static $Q \bar Q$ pair at infinite separation $U_{\infty}(T)$ \cite{Mocsy:2007jz}.
In Fig. \ref{fig:pot} we show the real part of the potential for several values of the temperature
Since on general grounds we expect $F_{\infty} \le V_{\infty}(T) \le U_{\infty}$ the above
choice of the potential provides the most binding potential compatible with the lattice data.
Following Ref. \cite{Mocsy:2007jz} we call this choice of Re$V(r,T)$ the maximally binding potential
(in Ref. \cite{Mocsy:2007jz} it was also labeled as set II).
Unfortunately lattice QCD cannot yet determine the imaginary part of the potential. Therefore
here we have to rely on perturbation theory. However, even in perturbation theory the form of
the imaginary part is only know for certain limiting cases corresponding to some hierarchy of
the relevant scales $1/r$, $T$, $m_D$ \cite{Brambilla:2008cx}. Therefore for the imaginary part we
choose the perturbative hard thermal loop (HTL) result obtained in Ref. \cite{Laine:2006ns} 
\begin{equation}
{\rm Im} V^{HTL}(r,T)=-\frac{i g^2 T C_F}{4 \pi}\phi(m_D r),
~\phi(x)=2 \int_0^{\infty} \frac{dz z}{(z^2+1)^2}
\left [ 1-\frac{sin(zx)}{zx} \right ],
\label{ImVLaine}
\end{equation}
which is in principle valid only for distances $r>1/m_D$ \cite{Brambilla:2008cx}.
However, the above expression has the nice feature that it vanishes for distances which are much
smaller than the inverse temperature. For the numerical values of $g^2(T)$ and $m_D(T)$ entering the
above formulas following Ref. \cite{Burnier:2007qm} we choose
\begin{equation}
g^2(T)=\frac{8 \pi^2}{9 \ln (9.082T/\Lambda_{\overline{MS}})},~~
m_D^2(T)=\frac{8 \pi^2}{3 \ln (7.547T/\Lambda_{\overline{MS}})},~\Lambda_{\overline{MS}}=300{\rm MeV}.
\label{parLaine}
\end{equation}
Due to the large numerical pref-actor in the argument of 
the logarithm the imaginary part of the potential is not too large
even at temperatures $T\simeq 200$MeV. Therefore we may consider this choice as a conservative
lower bound for it. The resulting Im$V(r,T)$ is shown in Fig. \ref{fig:pot}. 
In the next section we  present charmonium and bottomonium spectral functions for this choice
of the potential which is maximally binding and minimally dissipative. 
\begin{figure}
\includegraphics[width=8cm]{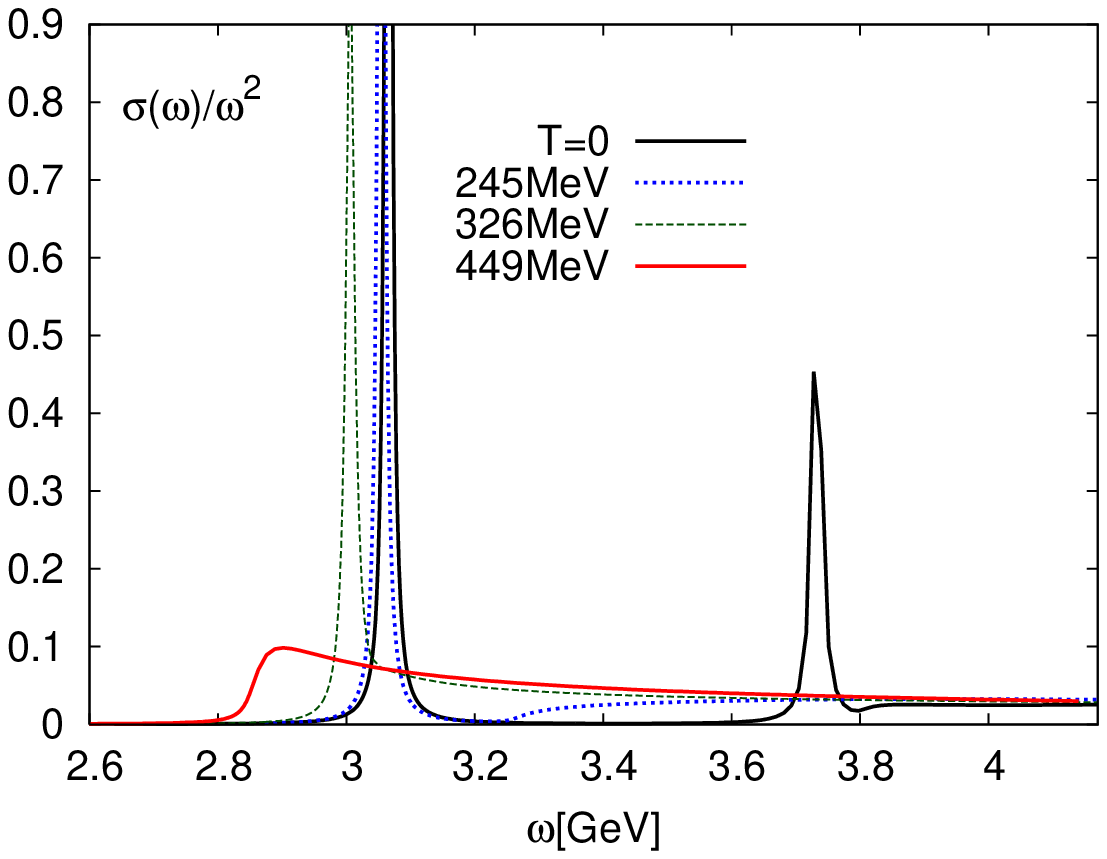}\hspace*{-0.3cm}
\includegraphics[width=8cm]{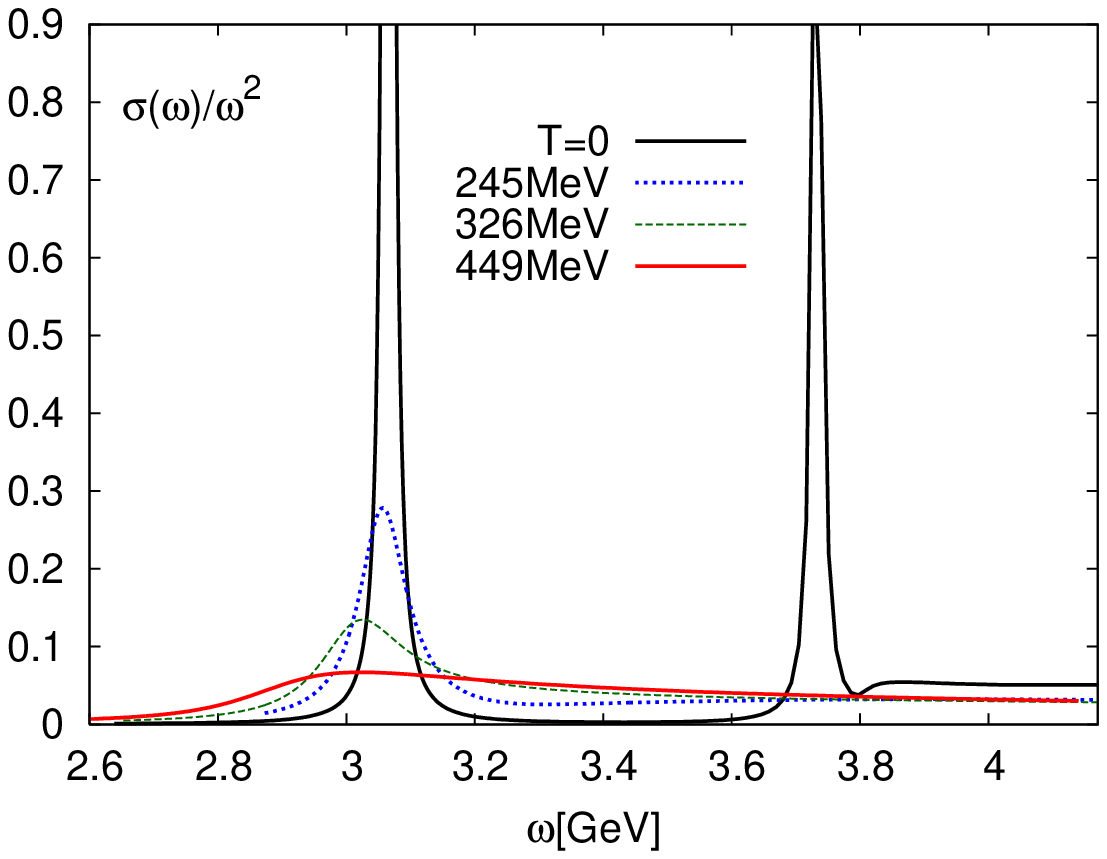}
\vspace*{-0.3cm}
\caption[]{The charmonium spectral functions function calculated in potential
model neglecting the imaginary part of the potential (left) and using both
the real and imaginary part of the potential (right).
The maximally binding potential has been used for the real part.
}
\label{fig:charm_spf}
\vspace*{-0.1cm}
\end{figure}

\section{Numerical results for the spectral functions}
\label{sec:num}
\begin{figure}[htb]
\includegraphics[width=8cm]{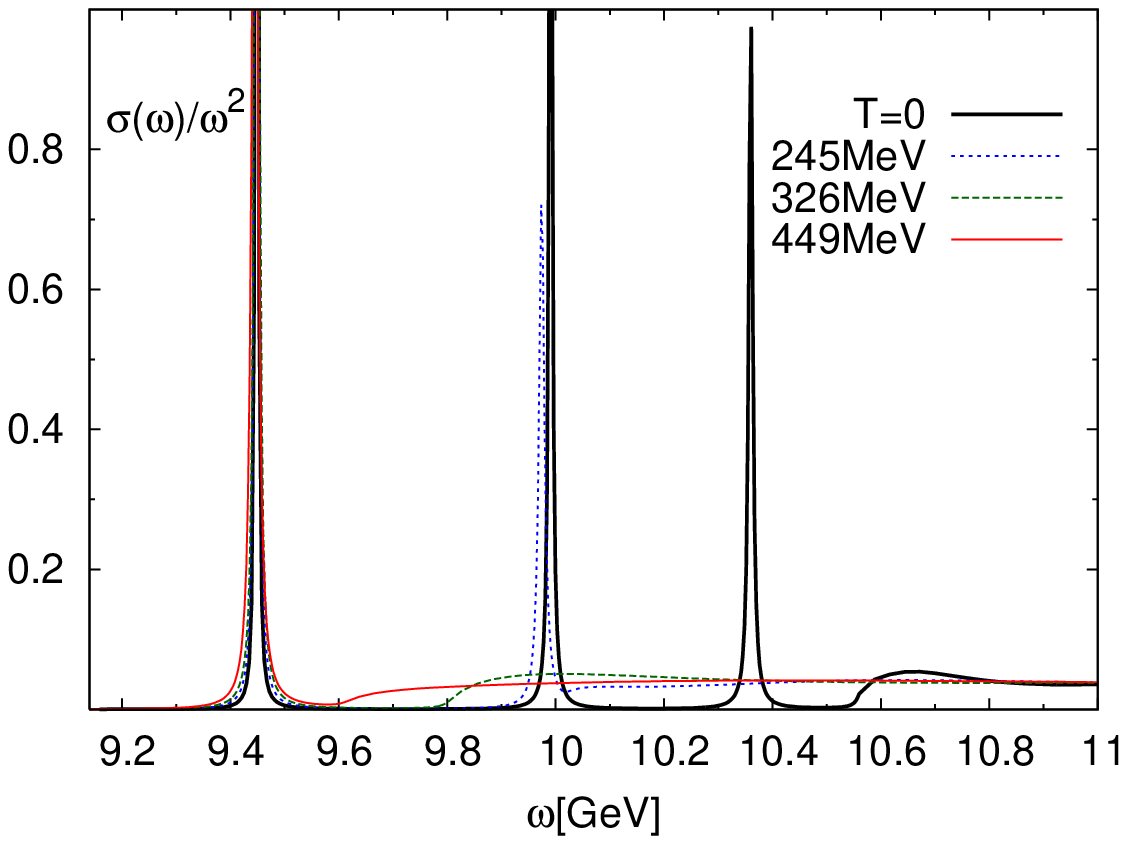}\hspace*{-0.3cm}
\includegraphics[width=8cm]{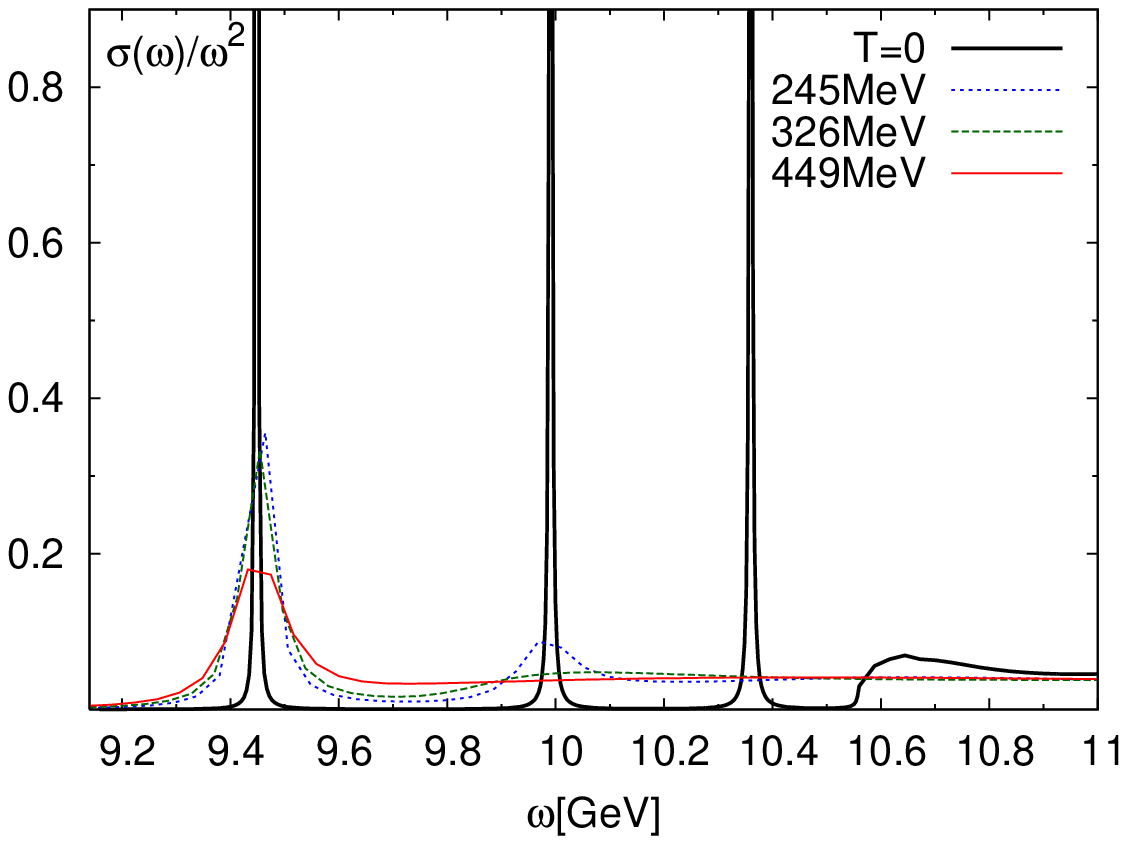}
\vspace*{-0.3cm}
\caption[]{The bottomonium spectral functions function calculated in potential
model neglecting the imaginary part of the potential (left) and using both
the real and imaginary part of the potential (right).
The maximally binding potential has been used for the real part.
}
\label{fig:bott_spf}
\vspace*{-0.1cm}
\end{figure}

With the potential discussed in the previous section we have solved the Schr\"odinger equation
for the non-relativistic Green functions and thus calculated the spectral function for S-wave
quarkonium. We used the numerical algorithm described in Ref. \cite{Burnier:2007qm}. 
First for a reference we need to calculate the zero temperature spectral function. At zero
temperature the potential has no imaginary part, however, Eq. (1) cannot be solved numerically
if the imaginary part of the potential is strictly zero. Therefore we need to introduce a small
imaginary part to the potential when solving the Schr\"odinger equation. In the past we used
a constant imaginary part equal to $0.01m$ for charmonium and $0.003m$ for bottomonium 
\cite{Mocsy:2007yj,Mocsy:2007jz}. Here
we use Im$V^{HTL}(r,T=148{\rm MeV})/10$ for the imaginary part. It is better than the previous choice
as it gives less artificial width for the ground state. Furthermore, we would like to incorporate
the physics of the string breaking into the zero temperature spectral functions. 
Following Ref. \cite{Mocsy:2007jz}
we do this by replacing the Coulomb plus linear form of the potential by a Yukawa form at $r>1.1$fm.
The parameters of the zero temperature potential can be found in \cite{Mocsy:2007jz}. For charm and bottom
quark mass we choose $m_c=1.29$GeV and $m_b=4.67$GeV respectively. This gives a reasonably good description
of the quarkonium spectrum at zero temperature as shown in Table \ref{tab:spect}.
\begin{table}[h]
\begin{center}
\begin{tabular}{|c|c|c|c|c|c|c|}
\hline
\multicolumn{3}{|c|}{Charmonia}        & & \multicolumn{3}{|c|}{Bottomonia}  \\
\hline
State        &  Model   & PDG     &&   State  &  Model       & PDG/LQCD \\
\hline
$1S$  &  3.067       & 3.068      &&   $1S$   &   9.451       &  9.443 \\
$2S$  &  3.720       & 3.674      &&   $2S$   &   9.988       & 10.016 \\
$1P$  &  3.536       & 3.525      &&   $3S$   &   10.357      & 10.355 \\
      &              &            &&   $1P$   &   9.864       &  9.900 \\
      &              &            &&   $2P$   &   10.247      & 10.260 \\
\hline
\end{tabular}
\caption{Spin-averaged charmonium and bottomonium masses in GeV at zero temperature calculated in our
  model and  compared to the experimental values \cite{pdg}. 
  To obtain the 2S spin averaged bottomonium mass we 
  used the lattice QCD value of $\eta_b(2S)$ mass from Ref. \cite{gray05}. }
\label{tab:spect}
\end{center}
\end{table} 
The $T=0$ spectral functions obtained using the above procedure are shown in Fig. \ref{fig:charm_spf}
and  Fig. \ref{fig:bott_spf} for charmonium and bottomonium respectively.

We would like to understand the effect of color screening, encoded in the real part
of the potential, as well as the  effects of dissipation, encoded in the imaginary part
of the potential on the quarkonium spectral functions. Therefore we first calculated the
quarkonium spectral functions neglecting the imaginary part of the potential. In terms
of the numerics this means that we used Im$V^{HTL}(r,T)/10$ in our calculations. The corresponding
charmonium and bottomonium spectral functions are shown in the left panels of 
Fig.  \ref{fig:charm_spf} and  Fig. \ref{fig:bott_spf} respectively. 
We see clear peak like structures in the spectral functions
which persist to temperature of $326$MeV for charmonium and $449$MeV for bottomonium. In the
case of bottomonium we also see a remnant of the $2S$ state in the spectral function at
$T=245$MeV. This is consistent with the results obtained in potential models, in particular,
with the calculation of Ref. \cite{Cabrera:2006wh} that used internal energy as a potential.

The situation changes dramatically when the imaginary part of the potential
${\rm Im} V(r,T)={\rm Im} V^{HTL}(r,T)$ is taken into account. The corresponding spectral
functions are also shown in the right panels of Fig. \ref{fig:charm_spf}
and  Fig. \ref{fig:bott_spf} for charmonium and bottomonium respectively. In the charmonium
spectral function we see a very broad peak at $T=245$MeV, which disappears at higher temperatures.
In the bottomonium spectral function the 2S peak disappears slightly above $T=245$MeV, while
the ground state peak becomes very broad at $T=450$MeV. 
These findings for the spectral functions
are consistent with upper bounds on the quarkonium dissociation temperatures 
obtained in Ref. \cite{Mocsy:2007jz}
by comparing the temperature dependence of the binding energies and the thermal widths, 
since the lattice
data used in both studies correspond to a deconfinement temperature of $T_{dec} \simeq T_c \simeq 204$MeV
\footnote{In Ref. \cite{Mocsy:2007jz} the dissociation temperatures were given in units of $T_c$.}.

\begin{figure}[htb]
\includegraphics[width=8cm]{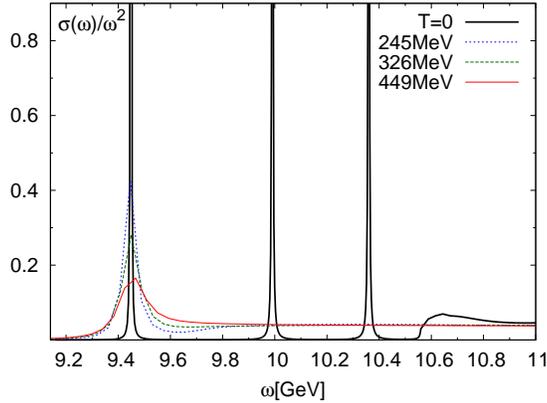}\hspace*{-0.3cm}
\vspace*{-0.3cm}
\caption[]{The bottomonium spectral functions function calculated using the complex potential
with real part corresponding to set I of Ref. \cite{Mocsy:2007jz}.
}
\label{fig:bot1_spf}
\vspace*{-0.1cm}
\end{figure}

We note that the choice of the real part of the potential used here 
gives the largest possible binding energies. Nevertheless, 
the binding energies
of all quarkonium states, except for 1S bottomonium are smaller than the temperature for $T\ge 245$MeV
and drop rapidly with increasing temperature \cite{Mocsy:2007jz}.
Thus, our approach of putting all medium effects into the potential should be at least marginally
correct. Binding energies are significantly smaller for other choices of the potential.
In particular, a theoretically motivated choice of the potential which is close to the free energy
was considered in Ref. \cite{Mocsy:2007jz}, where it was called set I. 
For this potential we see melting of
charmonium ground state at $T\simeq 250$ MeV even when no imaginary part is considered. At the same time 
the bottomonium spectral functions are not very different compared to 
the ones calculated with the maximally binding potential.
It is interesting to see how much our results change when switching the potential to set I and keeping
the imaginary part of the potential. The corresponding numerical data are shown in Fig. \ref{fig:bot1_spf}.
As one can see from the figure changing the real part of the potential has little 
effect on the bottomonium
spectral functions when the imaginary part is present. The dissolution of the excited states and the 
broadening of the ground state peak are mostly determined by the imaginary part. 
This also means that the microscopic mechanism behind the melting of charmonium and bottomonium
states could be quite different. For ground state charmonium screening effect play and important
role, while for the ground state bottomonium the melting will occur due to the large imaginary
part of the potential at high temperatures.

\section{Euclidean time correlation functions}
\label{sec:corr}
\begin{figure}
\includegraphics[width=8cm]{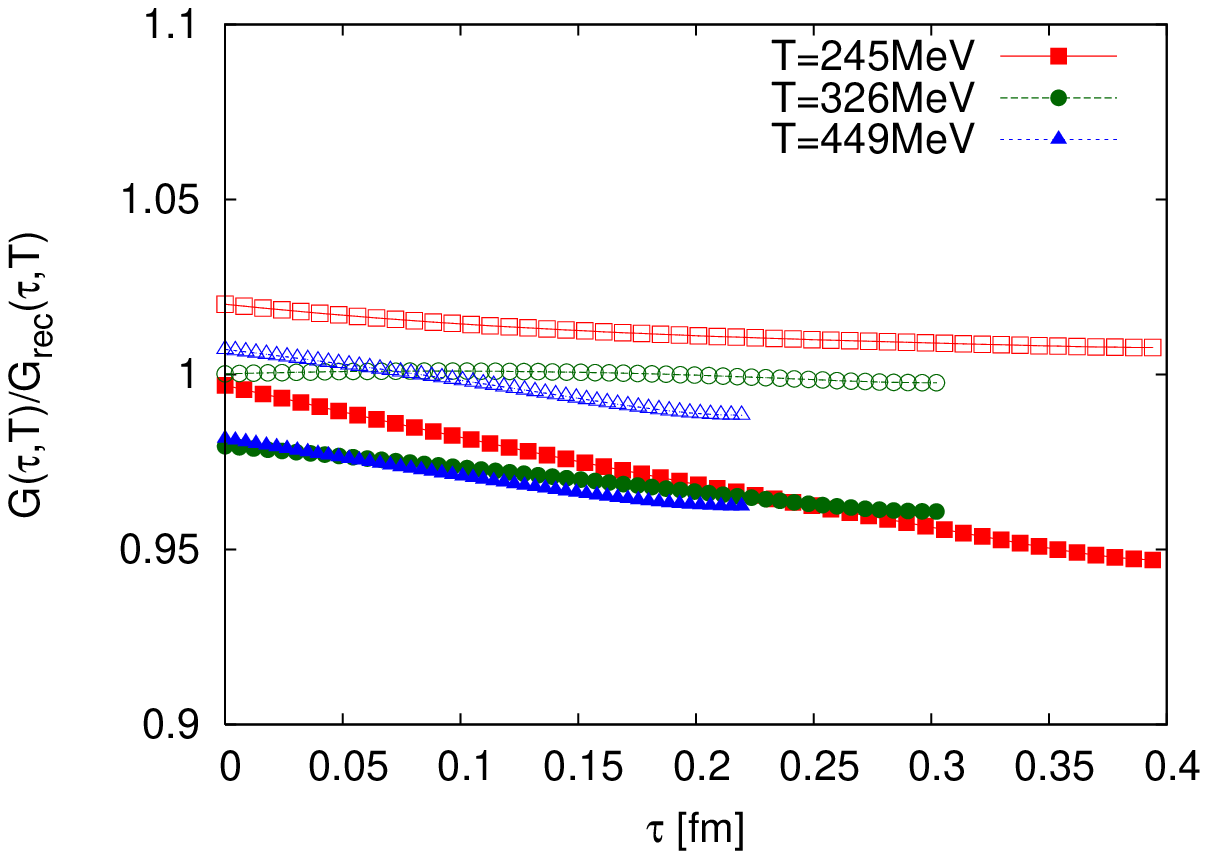}\hspace*{-0.3cm}
\includegraphics[width=8cm]{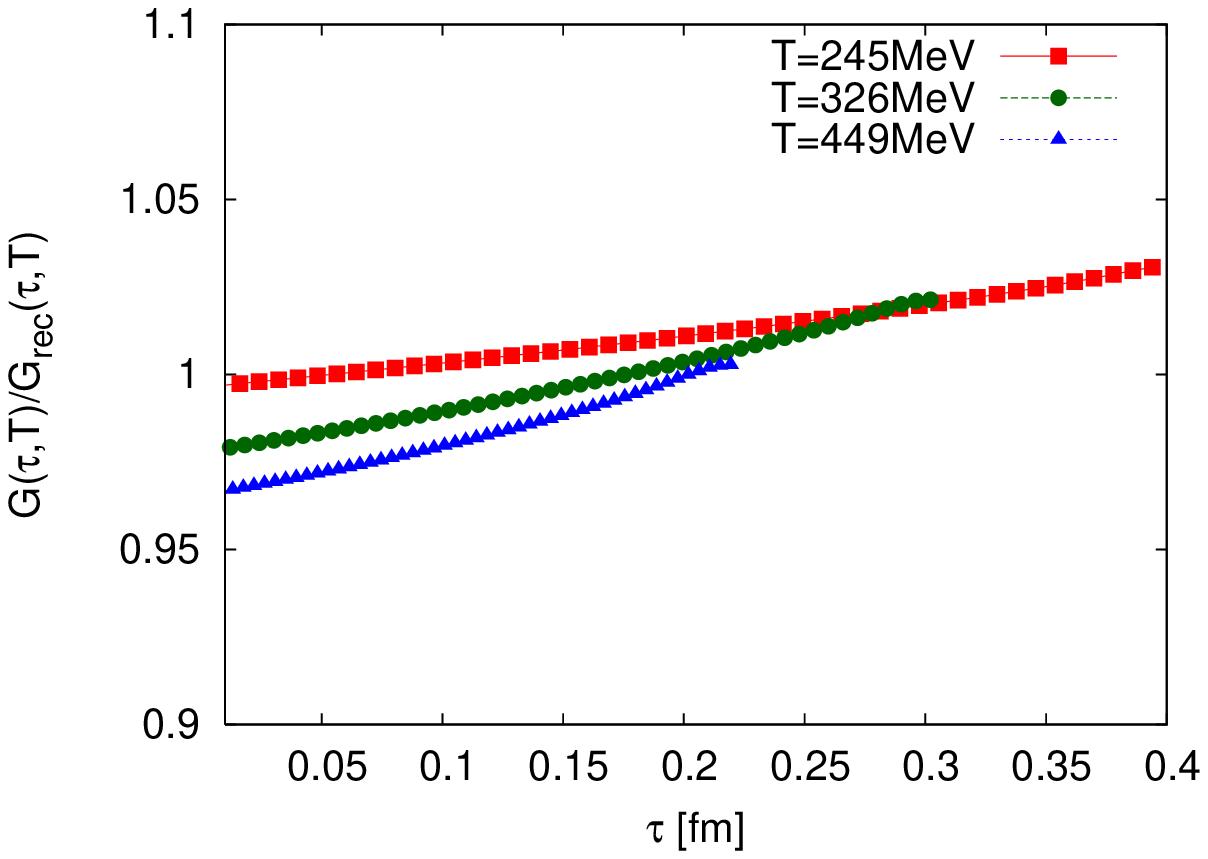}
\vspace*{-0.3cm}
\caption[]{The ratio $G(\tau,T)/G_{rec}(\tau,T)$ of charmonium (left) and bottomonium (right) correlators
for different temperatures calculated with complex potential. Open symbols
in the left plot correspond to the calculation where the imaginary part
of the potential was neglected.
}
\label{fig:rat}
\vspace*{-0.1cm}
\end{figure}

From the calculated quarkonium spectral functions it is easy to calculate the corresponding
Euclidean time correlation functions
\begin{equation}
G(\tau,T)=\int_0^{\infty} d \omega \sigma(\omega,T) K(\tau,\omega,T),~
K((\tau,\omega,T)=\frac{\cosh(\omega(\tau-1/(2T)))}{\sinh(\omega/(2T))}.
\end{equation}
It is important to compare the temperature dependence of the 
correlation functions obtained in potential model with the results of the lattice QCD
calculations. The temperature dependence of the correlation function on the lattice
is studied in terms of the ratio $G(\tau,T)/G_{rec}(\tau,T)$, where
\begin{equation}
G_{rec}(\tau,T)=\int_0^{\infty} d \omega \sigma(\omega,T=0) K(\tau,\omega,T).
\end{equation}
In the lattice calculations this ratio for pseudo-scalar channel that corresponds to
S-wave quarkonia, was found to be close to one 
\cite{Datta:2003ww,Datta:2006ua,Jakovac:2006sf,Petreczky:2008px}.
Potential model calculation with real potential could reproduce this feature. Therefore it
is interesting to see how close is $G(\tau,T)/G_{rec}(\tau,T)$ to unity when the imaginary
part of the potential is taken into account. The spectral functions calculated in the potential
model are not reliable far away from the threshold because of relativistic effects. Therefore
in Refs. \cite{Mocsy:2007yj,Mocsy:2007jz} the spectral functions calculated in the potential model
were smoothly matched to the perturbative result for the spectral functions at some higher energy
and the ratios $G(\tau,T)/G_{rec}(\tau,T)$ have been calculated from these matched spectral functions.
In this study we did not follow this procedure and only used the spectral functions calculated
in the potential model when constructing $G/(\tau,T)/G_{rec}(\tau,T)$. Our results for the ratio
of the correlators is shown in Fig. \ref{fig:rat}. In the case of the charmonium this ratio stays
close to unity even when the complex potential is used. However, $G(\tau,T)/G_{rec}(\tau,T)$ 
is noticeable smaller compared to the calculations
 where the imaginary part is neglected shown as open symbols in the
figure. In the bottomonium case the imaginary part of the potential had a significant effect and the ratio
$G(\tau,T)/G_{rec}(\tau,T)$  is no longer close to unity in odds with the available lattice data
\cite{Datta:2006ua,Jakovac:2006sf}. It is possible that this is due to fact that time scale relevant
for the ground state bottomonium is smaller than the time scales related to the medium and modeling
of the medium effects by a $T$-dependent potential is not accurate enough. 
However, the in-medium potential based description should work better for the excited bottomonium
states due to their smaller binding energies. 
Therefore we constructed
the ratio $G(\tau,T)/G_{rec}(\tau,T)$ assuming that the ground state bottomonoium is unmodified and
using only the spectral function above $\omega=9.7$GeV in the calculation of the correlation functions.
The corresponding results are also shown in Fig. \ref{fig:rat}. As one can see from the figure the
in-medium changes of the bottomonium spectral functions above $9.7$GeV, in particular, the meting of
excited bottomonium states do not produce any large change in the Euclidean correlation functions. 
The problem of bottomonium spectral functions, however, needs further investigations in light of 
recent attempts to calculate the properties
of ground state bottomonium at high temperatures using perturbative QCD \cite{Brambilla:2010vq}
and new lattice calculations bottomonium correlators in NRQCD at finite temperature \cite{Aarts:2010ek}. 
We also note that 
since we do not match the spectral functions to the perturbative form at higher energies the
calculated  $G(\tau,T)/G_{rec}(\tau,T)$ shows larger deviation from unity, especially at shorter
distance compared to previous studies \cite{Mocsy:2007yj,Mocsy:2007jz}.

\section{Conclusions}
\label{sec:concl}
In summary, we have calculated the quarkonium spectral functions in the deconfined region
using pNRQCD inspired potential model with complex potential. 
For the real part of the potential we considered the lattice QCD motivated maximum
binding potential first introduced in Ref. \cite{Mocsy:2007jz}, while for the imaginary part
we took the perturbative estimate from \cite{Laine:2006ns}. 
We find in particular that $J/\psi$ peak disappear for temperatures above $245$MeV and $\Upsilon(1S)$ peak
disappears for temperatures above $450$MeV. 
We found that dissipative effects encoded in the imaginary part of the potential play a crucial role in
the quarkonioum spectral functions and dissolution of quarkonium states. Similar conclusion have been
made very recently in the analysis of charmonium spectral functions using the T-matrix approach, where
dissipative effects are encoded in the heavy quark self energy \cite{Riek:2010py}.
This study also finds no bound state peaks in QGP.
In the future it would be interesting to study other choices for the imaginary part of the potential, which 
could lower the dissociation temperature for the ground state bottomonium as well as to study the 
corresponding Euclidean time correlation functions more in detail.

\section*{Acknowledgements}
This work has been supported  by contract DE-AC02-98CH10886 with the U.S. Department of Energy.
The numerical calculations of the quarkonium spectral functions has been performed and the algorithm
and the code described in Ref. \cite{Burnier:2007qm}. We thank Mikko Laine for providing this code
for us.

\end{document}